\documentclass[paper, nofootinbib]{revtex4}
\usepackage{amsfonts}
\usepackage{graphicx}
\usepackage[latin1]{inputenc}
\usepackage{amsmath}
\usepackage{amssymb}

\begin{document}
\title{Lagrangian for the Frenkel electron}

\author{Alexei A. Deriglazov}
\email{alexei.deriglazov@ufjf.edu.br}
\affiliation{%
Depto. de Matem\'atica, ICE, Universidade Federal de Juiz de Fora, MG, Brasil \\ and \\ Laboratory of Mathematical
Physics, Tomsk Polytechnic University, 634050 Tomsk, Lenin Ave. 30, Russian Federation}


%

\begin{abstract}
We found Lagrangian action which describes spinning particle on the base of non-Grassmann  vector and involves only one
auxiliary variable. It provides the right number of physical degrees of freedom and yields generalization of the
Frenkel and BMT equations to the case of an arbitrary electromagnetic field. For a particle with anomalous magnetic
moment, singularity in the relativistic equations generally occurs at the speed different from the speed of light.
Detailed discussion of the ultra-relativistic motion is presented in the work: A. A. Deriglazov and W. G. Ramirez,
World-line geometry probed by fast spinning particle, arXiv:1409.4756.
\end{abstract}

\maketitle 

\section{Introduction}\label{intr}
Relativistic description of rotational degrees of freedom of a body starting from the proper Lagrangian has a long
history, see \cite{corben:1968, hanson1974relativistic, mukunda1982, stepanov2012} and references therein. Since the
spin operators in quantum theory satisfy the angular-momentum algebra, closely related problem consist in establishing
of variational formulation which should lead to classical equations of spinning electron \cite{Frenkel, Frenkel2, BMT}.
One possibility here is to assume the Frenkel spin-tensor $J^{\mu\nu}$ be the composite quantity
$J^{\mu\nu}=2(\omega^\mu\pi^\nu-\omega^\nu\pi^\mu)$ formed by non-Grassmann vector $\omega^\mu$ and its conjugated
momentum $\pi^\nu$ \cite{grassberger1978, cognola1981lagrangian, mukunda1982, Alexei}. Since spin should be described
by two physical degrees of freedom, we need to impose some constraints on eight basic variables $\omega^\mu$ and
$\pi^\nu$. Inclusion of the constraints into a variational problem turns out to be rather nontrivial task. Though a
number of vector models \cite{balachandran:1980, bt, bbra, grassberger1978, cognola1981lagrangian} yield Frenkel and
BMT equations, they also contain extra degrees of freedom. At the classical level one can simply ignore them. However,
they should be taken into account during quantization procedure, this leads to quantum models essentially different
from the Dirac electron. In the recent works \cite{DPM1, DPM2, DPM3, DPW2} we partially solved this task by presenting
a number of equivalent Lagrangians with the right physical sector. Free theory can be described by the Lagrangian
without auxiliary variables\footnote{The last term in (\ref{i.1}) represents velocity-independent constraint which is
well known from classical mechanics. So, we might follow the classical-mechanics prescription to exclude the auxiliary
variable $g_4$ from the formulation. But this would lead to lose of manifest covariance of the formalism.}
\begin{eqnarray}\label{i.1}
L=-mc\sqrt{-\dot xN\dot x}+\sqrt{a_3}\sqrt{\dot\omega N\dot\omega}-\frac{1}{2}g_4(\omega^2-a_4),
\end{eqnarray}
where $N^{\mu\nu}\equiv\eta^{\mu\nu}-\frac{\omega^\mu\omega^\nu}{\omega^2}$ is projector on the plane transverse to the
direction of $\omega^\mu$. The free parameters $a_3$ and $a_4$ determine the value of spin. The corresponding
relativistic quantum mechanics has been identified \cite{DPM2} with one-particle (positive energy) sector of the Dirac
equation. The problem here is that even the minimal interaction $\frac{e}{c}A_\mu\dot x^\mu$ leads to a theory with the
number and algebraic structure of constraints different from those of free theory. So the interacting theory has been
constructed \cite{DPM3, DPW2} on the base of Lagrangian with four auxiliary variables $g_i$
\begin{eqnarray}\label{i.2}
L=\frac{1}{2(g_1g_3-g_7^2)}\left[ g_3\left(\dot xN\dot x\right)-2g_7\left(\dot xND\omega\right)+ g_1\left(D\omega
ND\omega\right) \right]+ \cr \frac{e}{c}A_\mu\dot
x^\mu-\frac{g_4}{2}(\omega^2-a_4)-\frac{g_1}{2}m^2c^2+\frac{g_3}{2}a_3 \,.
\end{eqnarray}
Spin interacts with $A^\mu$ through the derivative $D$ defined in Eq. (\ref{m.2}). This yields a generalization of
Frenkel and BMT equations to the case of an arbitrary electromagnetic field \cite{DPM3}. In the present work we obtain
more economic formulation which involves only one auxiliary variable.


We work in four-dimensional Minkowski space with the metric $\eta_{\mu\nu}=(-, +, +, +)$. For contraction of indexes we
use the notation $\dot x^\mu\dot x_\mu=\dot x^2$, ~ $\dot x^\mu N_{\mu\nu}\dot x^\nu=\dot x N\dot x$, ~
$N^\mu{}_\nu\dot x^\nu=(N\dot x)^\mu$, ~ $F^{\mu\nu}J_{\mu\nu}=(FJ)$, $F^{\mu\alpha}J_\alpha{}^\nu=(FJ)^{\mu\nu}$ and
so on.

\section{Lagrangian and Hamiltonian formulations}\label{cov}
Consider spinning particle with mass $m$, electric charge $e$ and magnetic moment $\mu$ interacting with an arbitrary
electromagnetic field
$F_{\mu\nu}=\partial_\mu A_\nu-\partial_\nu A_\mu=(F_{i0}=E_i, ~ F_{ij}=\epsilon_{ijk}B_k)$. 
We study the following Poincare and reparametrization invariant Lagrangian action on configuration space with
coordinates $x^\mu(\tau)$, $\omega^\mu(\tau)$ and $g(\tau)$:
\begin{eqnarray}\label{m.1}
S=\int d\tau\frac{1}{4g}\left[\dot xN\dot x+gD\omega ND\omega-\sqrt{\left[\dot xN\dot x+gD\omega
ND\omega\right]^2-4g(\dot xND\omega)^2}\right]- \cr \frac{g}{2}m^2c^2+\frac{\alpha}{2\omega^2}+\frac{e}{c}A\dot x.
\qquad \qquad \qquad \qquad \qquad
\end{eqnarray}
This depends on one free parameter $\alpha$ which determines spin of the particle. We take $\alpha=\frac{3\hbar^2}{4}$,
this corresponds to spin one-half particle, see below. The only auxiliary variable is $g$, this provides the mass-shell
condition (\ref{m.14.1}). It has been denoted
\begin{equation}\label{m1.1}
N^{\mu\nu}\equiv\eta^{\mu\nu}-\frac{\omega^\mu\omega^\nu}{\omega^2}\,, \quad \mbox{then} \quad N^{\mu\nu}\omega_\nu=0.
\end{equation}
Together with $\tilde N^{\mu\nu}\equiv\frac{\omega^\mu\omega^\nu}{\omega^2}$, this forms a pair of projectors:
$N+\tilde N=1$, ~ $N^2=N$, ~  $\tilde N^2=\tilde N$, ~ $N\tilde N=0$.
The square-root appeared in the Lagrangian seem to be typical structure \cite{hanson1974relativistic} for the models
which imply the Frenkel-type condition $J^{\mu\nu}{\cal P}_\nu=0$.

To introduce coupling of the position variable $x$ with electromagnetic field we have added the minimal interaction
$\frac{e}{c}A_\mu\dot x^\mu$. As for spin, it couples with $A^\mu$ through the term
\begin{eqnarray}\label{m.2}
D\omega^\mu\equiv\dot\omega^\mu-g\frac{e\mu}{c}F^{\mu\nu}\omega_\nu\,.
\end{eqnarray}
This is the only term we have found compatible with symmetries and constraints which should be presented in the theory.
In particular, under reparametrizations the variable $g$ transforms as $g=\frac{\partial\tau'}{\partial\tau}g'$. This
implies homogeneous transformation law of $D\omega$, $D\omega=\frac{\partial\tau'}{\partial\tau}D'\omega'$, and, at the
end, reparametrization invariance of the Lagrangian. In turn, this provides the expected mass-shell condition ${\cal
P}^2-\frac{e\mu}{2c}(FJ)+m^2c^2=0$, see below.

Switching off the spin variables $\omega^\mu$ from Eq. (\ref{m.1}), we arrive at familiar Lagrangian of spinless
particle $L=\frac{1}{2g}\dot x^2-\frac{g}{2}m^2c^2+\frac{e}{c}A\dot x$. Integrating over the auxiliary variable $g$ we
obtain
$L=-mc\sqrt{-\dot x^2}+\frac{e}{c}A\dot x$.
This is equivalent to the standard Lagrangian of spinless particle in terms of physical variables $\vec x(t)$,
$L=-mc\sqrt{c^2-\dot{\vec x}^2}+eA_0+\frac{e}{c}\vec A\dot{\vec x}$, if we restrict ourselves to the class of
increasing parameterizations of the world-line. This implies positive $g(\tau)$. So we study (\ref{m.1}) under the
assumptions
$\frac{dt}{d\tau}>0$, $g(\tau)>0$.
In the presence of spin, our Lagrangian is a complicated function of $g$ even in the case of free theory.
%
%

Let us construct Hamiltonian formulation of the model. Conjugate momenta for $x^\mu$, $\omega^\mu$ and $g$ are denoted
as $p^\mu$, $\pi^\mu$ and $\pi_g$. Besides, we use the condensed notation $\sqrt{\left[\dot xN\dot x+gD\omega
ND\omega\right]^2-4g(\dot xND\omega)^2}\equiv\sqrt{~ ~ }$ and ${\cal P}^\mu\equiv p^\mu-\frac{e}{c}A^\mu$. Contrary to
$p^\mu$, the canonical momentum ${\cal P}^\mu$ is $U(1)$ gauge-invariant quantity.

Since $\pi_g=\frac{\partial L}{\partial\dot g}=0$, the momentum $\pi_g$ represents the primary constraint, $\pi_g=0$.
Expressions for the remaining momenta, $p^\mu=\frac{\partial L}{\partial\dot x_\mu}$ and $\pi^\mu=\frac{\partial
L}{\partial\dot\omega_\mu}$, can be written in the form
\begin{eqnarray}\label{m.3}
{\cal P}^\mu=\frac{1}{2g}(N\dot x^\mu-K^\mu), \qquad K^\mu\equiv\frac{\left[\dot xN\dot x+gD\omega
ND\omega\right](N\dot x)^\mu-2g(\dot xND\omega)(ND\omega)^\mu}{\sqrt{~ ~ }};
\end{eqnarray}
\begin{eqnarray}\label{m.4}
\pi^\mu=\frac{1}{2}(ND\omega^\mu-R^\mu), \qquad R^\mu\equiv\frac{\left[\dot xN\dot x+gD\omega
ND\omega\right](ND\omega)^\mu-2(\dot xND\omega)(N\dot x)^\mu}{\sqrt{~ ~ }}.
\end{eqnarray}
The functions $K^\mu$ and $R^\mu$ obey the following remarkable identities
\begin{eqnarray}\label{m.5}
K^2=\dot xN\dot x, \quad R^2=D\omega ND\omega, \quad KR=-\dot x ND\omega, \cr \dot xR+D\omega K=0, \quad \dot
xK+gD\omega R=\sqrt{~ ~ } \,.
\end{eqnarray}
Due to Eq. (\ref{m1.1}), contractions of the momenta with $\omega^\mu$ vanish, that is we have the primary constraints
$\omega\pi=0$ and ${\cal P}\omega=0$. One more primary constraint, ${\cal P}\pi=0$, is implied by (\ref{m.5}).

Hence we deal with a theory with four primary constraints. Hamiltonian has the form
\begin{eqnarray}\label{m.6}
H=p\dot x+\pi\dot\omega-L+\lambda_iT_i,
\end{eqnarray}
where $\lambda_i$ are the Lagrangian multipliers for the primary constraints $T_i$. To construct manifest form of the
Hamiltonian, we note the equalities
${\cal P}^2=\frac{1}{2g^2}[\dot xN\dot x-\dot xK]$ and $\pi^2=\frac{1}{2}[D\omega ND\omega-D\omega R]$.
Then, using (\ref{m.5}) we obtain
\begin{eqnarray}\label{m.8}
\frac{g}{2}{\cal P}^2+\frac{1}{2}\pi^2=L_0,
\end{eqnarray}
where $L_0$ is the first line in Eq. (\ref{m.1}). Further, using Eqs. (\ref{m.5}) we have
\begin{eqnarray}\label{m.9}
p\dot x+\pi\dot\omega\equiv{\cal P}\dot x+\frac{e}{c}A\dot x+\pi D\omega+g\frac{e\mu}{c}(\pi
F\omega)=2L_0+\frac{e}{c}A\dot x-g\frac{e\mu}{4c}(FJ),
\end{eqnarray}
where the Frenkel-type spin-tensor appeared
\begin{eqnarray}\label{m.10}
J^{\mu\nu}=2(\omega^\mu\pi^\nu-\omega^\nu\pi^\mu).
\end{eqnarray}
Using (\ref{m.9}) and (\ref{m.8}) in (\ref{m.6}) we arrive at the Hamiltonian
\begin{eqnarray}\label{m.11}
H=\frac{g}{2}\left({\cal P}^2-\frac{e\mu}{2c}(FJ)+m^2c^2\right)+\frac12\left(\pi^2-\frac{\alpha}{\omega^2}\right)+ \cr
\lambda_5(\omega\pi)+ \lambda_6({\cal P}\omega)+\lambda_7({\cal P}\pi)+\lambda_{g}\pi_g.
\end{eqnarray}
The fundamental Poisson brackets $\{x^\mu, p^\nu\}=\eta^{\mu\nu}$ and $\{\omega^\mu, \pi^\nu\}=\eta^{\mu\nu}$ imply
\begin{eqnarray}\label{m.12}
\{x^\mu, {\cal P}^\nu\}=\eta^{\mu\nu}, \quad \{{\cal P}^\mu, {\cal P}^\nu\}=\frac{e}{c}F^{\mu\nu},
\end{eqnarray}
\begin{eqnarray}\label{m.13}
\{J^{\mu\nu},J^{\alpha\beta}\}= 2(\eta^{\mu\alpha} J^{\nu\beta}-\eta^{\mu\beta} J^{\nu\alpha}-\eta^{\nu\alpha}
J^{\mu\beta} +\eta^{\nu\beta} J^{\mu\alpha})\,.
\end{eqnarray}
According to Eq. (\ref{m.13}) the spin-tensor is generator of Lorentz algebra $SO(1,3)$. As $\omega\pi$, $\omega^2$ and
$\pi^2$ are Lorentz-invariants, they have vanishing Poisson brackets with $J^{\mu\nu}$. To reveal the higher-stage
constraints we write the equations $\dot T_i=\{T_i, H\}=0$. The Dirac procedure stops on  third stage with the
following equations
\begin{eqnarray}\label{m.14}
\pi_g=0 ~ ~ \quad &\Rightarrow& \quad  {\cal P}^2-\frac{e\mu}{2c}(FJ)+m^2c^2=0 \quad
\Rightarrow \quad \lambda_6C+\lambda_7D=0\,,\label{m.14.1} \\
(\omega\pi)=0 ~ ~ \quad &\Rightarrow& \quad ~  \pi^2-\frac{\alpha}{\omega^2}=0\,,\label{m.14.2} \\
({\cal P}\omega)=0 ~ \quad  &\Rightarrow & \quad ~  \lambda_7=\frac{gC}{M^2c^2}\,,\label{m.14.3} \\
({\cal P}\pi)=0 ~ \quad &\Rightarrow&\quad ~  \lambda_6=-\frac{gD}{M^2c^2}\,.\label{m.14.4} 
\end{eqnarray}
We have denoted
\begin{eqnarray}\label{m.15}
M^2=m^2-\frac{e(2\mu+1)}{4c^3}(FJ), \quad
C=-\frac{e(\mu-1)}{c}(\omega F{\cal P})+\frac{e\mu}{4c}(\omega\partial)(FJ), \cr D=-\frac{e(\mu-1)}{c}(\pi F{\cal
P})+\frac{e\mu}{4c}(\pi\partial)(FJ).
\end{eqnarray}
The last equation from (\ref{m.14.1}) turns out to be a consequence of (\ref{m.14.3}) and (\ref{m.14.4}) and can be
omitted. Hence the Dirac procedure revealed two secondary constraints written in Eqs. (\ref{m.14.1}) and
(\ref{m.14.2}), and fixed the Lagrangian multipliers $\lambda_6$ and $\lambda_7$. The multipliers $\lambda_g$ and
$\lambda_5$ and the auxiliary variable $g$ have not been determined. $H$ vanishes on the complete constraint surface,
as it should be in a reparametrization-invariant theory.

We summarized the algebra of Poisson brackets between constraints in the Table \ref{tabular:monster-algebra1}.
\begin{table}
\begin{center}
\begin{tabular}{c|c|c|c|c|c}
                              & $\qquad T_1 \qquad$  & $T_3$                   & $T_5$                 & $T_6$    & $T_7$     \\  \hline \hline
$T_1=\mathcal{P}^2- $         & 0             & 0             & 0                                   & -2C   & -2D     \\
$\frac{\mu e}{2c}(FJ)+m^2c^2$
                              &               &               &           &                           &             \\
\hline
$T_3=\pi^2-\frac{\alpha}{\omega^2}$               & 0             & 0                & $-2T_3\approx 0$          & $-2T_7\approx 0$   & $\frac{2\alpha}{(\omega^2)^2}T_6\approx 0$\\
& ${}$ &      &           &
&        \\
\hline
$T_5=\omega\pi$               & 0     & $2T_3\approx 0$       &0     & $-T_6\approx 0$     & $T_7\approx 0$\\
& ${}$ &      &           &
&        \\
\hline
$T_6={\cal P}\omega$       &$2C$  &$2T_7\approx 0$       &$T_6\approx 0$            & 0       &$T_1-M^2c^2$  \\
 &               &               &           &                                                                               &$\approx -M^2c^2$ \\
\hline
$T_7={\cal P}\pi$          & $2D$      &$-\frac{2\alpha}{(\omega^2)^2}T_6\approx 0$           &$-T_7\approx 0$          &$-T_1+M^2c^2$  &0\\
&    &       &                                                                                                                                      &$\approx M^2c^2$  &\\
\hline
\end{tabular}
\end{center}
\caption{Algebra of constraints.} \label{tabular:monster-algebra1}
\end{table}
The constraints $\pi_g$, $T_1$, $T_3$ and $T_5$ form the first-class subset, while $T_6$ and $T_7$ represent a pair of
second class.  The presence of two primary first-class constraints $\pi_g$ and $T_5$ is in correspondence with the fact
that two lagrangian multipliers remain undetermined within the Dirac procedure. This also indicates on two local
symmetries which must be presented in the theory. One of them is the standard reparametrization invariance. Another is
the spin-plane symmetry discussed in the next section.

%
%

Hamiltonian (\ref{m.11}) determines evolution of the basic variables according the following equations
\begin{eqnarray}\label{uf4.5}
\dot x^\mu=g(T^{\mu}{}_\nu{\cal P}^\nu+Y^\mu), \quad \qquad \dot{\cal P}^\mu=\frac{e}{c}(F\dot x)^\mu+g\frac{\mu
e}{4c}\partial^\mu(FJ),
\end{eqnarray}
\begin{eqnarray}\label{uf4.6}
\dot\omega^\mu=g\frac{e\mu}{c}(F\omega)^\mu+g\frac{C}{M^2c^2}{\cal P}^\mu+\pi^\mu+\lambda_5\omega^\mu, \cr
\dot\pi^\mu=g\frac{e\mu}{c}(F\pi)^\mu+g\frac{D}{M^2c^2}{\cal
P}^\mu-\frac{\alpha}{(\omega^2)^2}\omega^\mu-\lambda_5\pi^\mu,
\end{eqnarray}
We have denoted
\begin{eqnarray}\label{uf4.71}
T^{\mu\nu}=\eta^{\mu\nu}-(\mu-1)a(JF)^{\mu\nu}, \qquad Y^\mu=\frac{\mu a}{4}J^{\mu\alpha}\partial_\alpha(JF)\,, \cr
a=\frac{e}{2M^2c^3}=\frac{-2e}{4m^2c^3-e(2\mu+1)(JF)}.
\end{eqnarray}
The interaction leads to modification of the Lorentz-force equation. Only for the ``classical'' value of magnetic
moment, $\mu=1$, and constant electromagnetic field the constraints (\ref{m.14.3}) and (\ref{m.14.4}) would be the same
as in the free theory, $\lambda_6=\lambda_7=0$. Then $T^{\mu\nu}=\eta^{\mu\nu}$, $Y^{\mu}=0$, and four-velocity becomes
proportional to ${\cal P}^\mu$, see (\ref{uf4.5}). Contribution of anomalous magnetic moment $\mu\ne 1$ to the
difference between $\dot x^\mu$ and ${\cal P}^\mu$ is proportional to $\frac{J}{c^3}\sim\frac{\hbar}{c^3}$, while the
term with a gradient of field is proportional to $\frac{J^2}{c^3}\sim\frac{\hbar^2}{c^3}$.

All the basic variables have ambiguous evolution. $x^\mu$ and ${\cal P}^\mu$ have one-parametric ambiguity due to $g$
while $\omega$ and $\pi$ have two-parametric ambiguity due to $g$ and $\lambda_5$. The quantities $x^\mu$, ${\cal
P}^\mu$ and the spin-tensor $J^{\mu\nu}$ turn out to be invariant under spin-plane symmetry. So they can be observable
quantities. Equations (\ref{uf4.5}) together with
\begin{eqnarray}\label{uf4.9}
\dot J^{\mu\nu}&=& g\frac{e\mu}{c}(FJ)^{[\mu\nu]}+2{\cal P}^{[\mu}\dot x^{\nu]}\,,
\end{eqnarray}
form a closed system.  The remaining ambiguity due to $g$ presented in these equations reflects the reparametrization
symmetry of the theory.

The term $\frac{\alpha}{2\omega^2}$ in the Lagrangian (\ref{m.1}) provides the constraint
$\omega^2\pi^2=\alpha=\frac{3\hbar^2}{4}$. Together with $\omega\pi=0$, this implies fixed value of spin
\begin{eqnarray}\label{uf4.12.1}
J^{\mu\nu}J_{\mu\nu}=8(\omega^2\pi^2-(\omega\pi)^2)=6\hbar^2\,,
\end{eqnarray}
for any solution to the equations of motion. Besides, the constraints $\omega{\cal P}=\pi{\cal P}=0$ imply the Pirani
condition \cite{pirani:1956, dixon:1964, tulczyjew:1959}
\begin{eqnarray}\label{uf4.12.2}
J^{\mu\nu}{\cal P_\nu}=0\,.
\end{eqnarray}

The variables $x$, ${\cal P}$ and $J$ have vanishing Poisson brackets with the second and third terms in (\ref{m.11}).
Hence these terms do not contribute into equations (\ref{uf4.5}) and (\ref{uf4.9}), and can be omitted from
Hamiltonian. Further, we can construct the Dirac bracket for the second-class pair $T_6$ and $T_7$, after that they
also can be omitted from (\ref{m.11}). Then the relativistic Hamiltonian acquires an expected form
\begin{eqnarray}\label{uf4.12}
H=\frac{g}{2}\left({\cal P}^2-\frac{e\mu}{2c}(FJ)+m^2c^2\right).
\end{eqnarray}
The equations (\ref{uf4.5}) and (\ref{uf4.9}) follow from this $H$ with use of Dirac bracket, $\dot z=\{z, H\}_{DB}$.

The first equation from (\ref{uf4.5}) together with $T_1$\,-constraint can be used to exclude the variables ${\cal
P}^\mu$ and $g$. For $g$ we obtain $\frac{\sqrt{-g_{\mu\nu}\dot x^\mu\dot x^\nu}}{m_rc}$, where the effective metric
$g_{\mu\nu}$ is given by (\ref{L.13}). So, the presence of $g$ in Eq. (\ref{m.2}) implies highly non-linear interaction
of spinning particle with electromagnetic field. Excluding ${\cal P}^\mu$ and $g$, we obtain closed system of
Lagrangian equations for the set $x, J$
\begin{eqnarray}\label{L.7}
D_g\left[m_r(\tilde T D_gx)^\mu\right]=\frac{e}{c^ 2}(FD_gx)^\mu+ \frac{\mu
e}{4m_rc^3}\partial^\mu(JF)+D_g(\frac{b}{ac}Y^\mu)\,,
\end{eqnarray}
\begin{eqnarray}\label{L.8}
\dot J^{\mu\nu}=\frac{e\mu\sqrt{-\dot xg\dot x}}{m_rc^2}(FJ)^{[\mu\nu]}-
\frac{2b(\mu-1)m_rc}{\sqrt{-\dot xg\dot x}}\dot x^{[\mu}(JF\dot x)^{\nu]}+\frac{2b}{a}\dot x^{[\mu}Y^{\nu]}\,.
\end{eqnarray}
Besides, all solutions satisfy the Lagrangian analog of Pirani condition
\begin{eqnarray}\label{L.9}
J^{\mu\nu}[(\tilde T\dot x)_\nu-\frac{b}{am_rc}\sqrt{-\dot xg\dot x}Y_\nu]=0,
\end{eqnarray}
as well as to the value-of-spin condition $J^{\mu\nu}J_{\mu\nu}=6\hbar^2$. We have denoted by
\begin{eqnarray}\label{pp7}
\tilde T^{\mu\nu}=\eta^{\mu\nu}+(\mu-1)b(JF)^{\mu\nu}, \qquad 
b=\frac{2a}{2+(\mu-1)a(JF)}\equiv\frac{-2e}{4m^2c^3-3e\mu(JF)},
\end{eqnarray}
the inverse matrix for $T$,  Eq. (\ref{uf4.71}).
Interaction of spin with the external field yields the radiation mass $m_r$
\begin{eqnarray}\label{L.11}
m_r^2=m^2-\frac{\mu e}{2c^3}(JF)-\frac{(YgY)}{c^2}\,,
\end{eqnarray}
as well as the effective metric
\begin{eqnarray}\label{L.13}
g_{\mu\nu}=(\tilde T^T\tilde T)_{\mu\nu}=[\eta+b(\mu-1)(JF+FJ)+b^2(\mu-1)^2FJJF]_{\mu\nu}, \cr
D_g=\frac{1}{\sqrt{-\dot xg\dot x}}\frac{d}{d\tau}\,. \qquad \qquad \qquad \qquad \qquad \quad
\end{eqnarray}
The equations (\ref{L.7})-(\ref{L.9}) coincide with those obtained in \cite{DPM3} from the Lagrangian with four
auxiliary variables. In the approximation $O^3(J, F, \partial F)$ and when $\mu=1$ they coincide with Frenkel
equations.

Let us specify the equation for spin precession to the case of uniform and stationary field, supposing also $\mu=1$ and
taking physical time as the parameter, $\tau=t$. Then (\ref{L.9}) reduces to the Frenkel condition, $J^{\mu\nu}\dot
x_\nu=0$, while (\ref{L.8}) reads $\dot J^{\mu\nu}=\frac{e\sqrt{-\dot x^2}}{m_rc^2}(FJ)^{[\mu\nu]}$. We decompose
spin-tensor on electric dipole moment $\vec D$ and Frenkel spin-vector $\vec S$ as follows:
\begin{eqnarray}\label{L.14.6}
J^{\mu\nu}=(J^{i0}=D^i, ~ J^{ij}=2\epsilon_{ijk}S_k).
\end{eqnarray}
Then $\vec D=-\frac{2}{c}\vec S\times\vec v$, while precession of $\vec S$ is given by
\begin{eqnarray}\label{L.14.5}
\frac{d\vec S}{dt}=\frac{e\sqrt{c^2-\vec v^2}}{m_rc^3}\left[-\vec E\times(\vec v\times\vec S)+c\vec S\times\vec
B\right].
\end{eqnarray}

\section{Spin surface and associated spin fiber bundle $\mathbb{T}^4$.}\label{sfb}

While spin-sector of our model consists of the basic variables $\omega^\mu$ and $\pi^\mu$, quantum mechanics obtained
in terms of spin-tensor $J^{\mu\nu}$. The passage from $\omega$ and $\pi$ to $J$ is not a change of variables, and
acquires a natural interpretation in the geometric construction described below. Generalization of this construction on
the case of $SO(k, n)$ Lie-Poisson manifold can be found in \cite{deriglazov2012variational}.

In the previous section we have obtained the following constraints in spin-sector:
\begin{eqnarray}\label{uf4.14}
{\cal P}\omega=0,  \qquad {\cal P}\pi=0,
\end{eqnarray}
\begin{eqnarray}\label{uf4.14.1}
\omega\pi=0, \qquad  \pi^2-\frac{\alpha}{\omega^2}=0,
\end{eqnarray}
It should be noticed that the Lagrangian (\ref{i.2}) implies $\omega\pi=0$, $\pi^2-a_3=0$ and $\omega^2-a_4=0$ instead
of (\ref{uf4.14.1}). So, the Lagrangian (\ref{m.1}) does not appear from (\ref{i.2}) by removing the auxiliary
variables $g_3$, $g_4$ and $g_7$.

To see the meaning of Lorentz-invariant constraints (\ref{uf4.14}) and (\ref{uf4.14.1}), we consider this surface in
Lorentz frame which implies ${\cal P}^\mu=({\cal P}^0, \vec 0)$. Then Eqs. (\ref{uf4.14}) mean $\omega^0=\pi^0=0$.
Taking this into account, the constraints (\ref{uf4.14.1}) determines the following surface in $\mathbb
{R}^6(\vec\omega, \vec\pi)$
\begin{eqnarray}\label{uf4.15}
\mathbb{T}^4=\{~ \vec\omega\vec\pi=0, \quad \vec\pi^2-\frac{\alpha}{\vec\omega^2}=0 ~ \},
\end{eqnarray}
that is $\vec\omega$ and $\vec\pi$ represent a pair of orthogonal vectors with ends attached to the hyperbole
$y=\frac{\alpha}{x}$. The constraints  (\ref{uf4.14}) imply $J^{\mu\nu}{\cal P}_\nu=0$. In the rest frame this gives
$J^{i0}=0$, that is the spin-tensor has only three components which we identify with non-relativistic spin-vector,
$J_{ij}=2\epsilon_{ijk}S_k$. Due to the constraints (\ref{uf4.15})  the spin-vector belong to two-dimensional sphere of
radius $\sqrt{\alpha}$
\begin{eqnarray}\label{uf4.16}
J_{ij}J_{ij}= 8\alpha, \quad \mbox{or} \quad \vec S^2=\alpha, \quad \mbox{so we assume} \quad
\alpha=\frac{3\hbar^2}{4}.
\end{eqnarray}
We call this the spin surface. The chosen value of parameter corresponds to spin one-half particle.

Hence, to describe spin in the rest frame, we have six-dimensional space of basic variables $\mathbb {R}^6(\vec\omega,
\vec\pi)$, the spin-tensor space $\mathbb {R}^3(J_{ij}\sim\vec S)$ and the map
\begin{eqnarray}\label{uf4.17}
f: ~ \mathbb{R}^{6} ~ \rightarrow ~ \mathbb{R}^{3}, \quad f: (\vec\omega, \vec\pi) ~ \rightarrow ~ \vec
S=\vec\omega\times\vec\pi, \quad \mbox{rank}\frac{\partial(S_i)}{\partial(\omega_j, \pi_k)}=3.
\end{eqnarray}
According to previous section, all trajectories $\vec{\omega}(\tau), \vec{\pi}(\tau)$ lie in the manifold
(\ref{uf4.15}) of $\mathbb{R}^{6}$. $f$ maps the manifold $\mathbb{T}^4$ onto spin surface,
$f(\mathbb{T}^4)=\mathbb{S}^2$.

Denote $\mathbb{F}^2_S\in\mathbb{T}^4$ preimage of a point ${\vec{S}}\in\mathbb{S}^2$,
$\mathbb{F}^2_S=f^{-1}({\vec{S}})$. Let $(\vec{\omega}, \vec{\pi})\in\mathbb{F}^2_S$. Then the two-dimensional manifold
$\mathbb{F}^2_S$ consist of the pairs $(k\vec\omega, \frac{1}{k}\vec\pi)$, $k\in\mathbb{R^{+}}$, as well as those
obtained by rotation of $(k\vec\omega, \frac{1}{k}\vec\pi)$ in the plane of vectors $\vec{\omega}$ and $\vec{\pi}$. So
elements of $\mathbb{F}^2_S$ are related by two-parametric transformations
\begin{eqnarray}\label{uf4.18}
\vec{\omega}'=\vec{\omega}k\cos\beta+\vec{\pi}\frac{k|\vec\omega|}{|\vec\pi|}\sin\beta, \qquad
\vec{\pi}'=-\vec{\omega}\frac{|\vec\pi|}{k|\vec\omega|}\sin\beta+\vec{\pi}\frac{1}{k}\cos\beta.
\end{eqnarray}
In the result, the manifold $\mathbb{T}^4$ acquires natural structure of fiber bundle
$\mathbb{T}^4=(\mathbb{S}^2, \mathbb{F}^2, f)$
with base $\mathbb{S}^2$, standard fiber $\mathbb{F}^2$, projection map $f$ and structure group given by
transformations (\ref{uf4.18}). As local coordinates of $\mathbb{T}^4$ adjusted with the structure of fiber bundle we
can take $k, \beta$, and two coordinates of the vector $\vec S$. By construction, the structure-group transformations
leave inert points of base, $\delta S_i=0$.

The Lorentz-invariant equations (\ref{uf4.14}), (\ref{uf4.14.1}) together with the map
$J^{\mu\nu}=2\omega^{[\mu}\pi^{\nu]}$ represent this construction in an arbitrary Lorentz frame. In the dynamical
realization given in previous section, structure group acts independently at each instance of time and turn into the
local symmetry. $k$\,-transformations provide reparametrization invariance of the action (\ref{m.1}). The spin-plane
rotations $\beta$ are associated with the first-class constraints $T_3$ and $T_5$ and selects $J$ as the physical
(observable) variable.

\section{Discussion}\label{cov3}
We obtained the generalization (\ref{L.7}) and (\ref{L.8}) of Frenkel and BMT equations to the case of an arbitrary
electromagnetic field. They follow from the Lagrangian (\ref{m.1}) which also yields the constraints (\ref{m.14}),
(\ref{uf4.12.1}) and (\ref{L.9}), providing the right number of physical degrees of freedom. Some relevant comments are
in order.

The relativistic equation (\ref{L.14.5}) automatically incorporates the Thomas precession \cite{Thomas, weinbergGC,
stepanov2012, Rebilas}. Indeed, let in instantaneous rest frame of the particle we have $F'^{\mu\nu}=(\vec
E'=\mbox{const}, ~ \vec B'=0)$. Then Eq. (\ref{L.14.5}) tell us that spin does not experience a torque in the rest
frame, $\frac{d\vec S'}{dt'}=0$. Consider a frame where the particle has velocity $\vec v$. In this frame the field is
$F^{\mu\nu}=(\vec E, ~  \vec B=\frac{1}{c}\vec v\times\vec E)$, where $\vec E$ is determined by Lorentz boost of $\vec
E'$ \cite{weinbergGC}. An observer in the laboratory frame detects the Thomas precession  (\ref{L.14.5}). Expressing
$\vec B$ through $\vec E$, the equation  (\ref{L.14.5}) can be written as follows: $\frac{d\vec
S}{dt}=\frac{e\sqrt{c^2-\vec v^2}}{m_rc^3}\vec v\times(\vec S\times\vec E)$.

Classical analog of the Pauli Hamiltonian \cite{foldy:1978} contains the term
$\frac12\vec S\cdot\vec E\times\vec v+c\vec S\cdot\vec B$,
while the relativistic theory (\ref{uf4.12}) implies
$\frac{c}{4}J^{\mu\nu}F_{\mu\nu}=\vec S\cdot\vec E\times\vec v+c\vec S\cdot\vec B$.
Both Hamiltonians are written in a laboratory system. The difference is the famous one-half factor. Our analysis
clearly shows the origin of this discrepancy on the classical level: we deal with two different sets of variables. Our
variables obey noncommutative Dirac brackets while in the Pauli theory the brackets supposed to be canonical. To
compare the Hamiltonians, we need manifest form of (time-dependent) canonical transformation among the two
formulations. Probably, the projection operator method for diagonalization of Dirac brackets \cite{Nakamura1993,
Nakamura2011, Nakamura2014} could be used to this aim.

Even for uniform fields, behavior of our spinning particle with anomalous magnetic moment ($\mu\ne 1$) differs from
that of Frenkel and BMT. This is due to two structural modifications implied by the Lagrangian which provides the
necessary constraints\footnote{Comparing with Frenkel, our formulation fixes the value of spin.}. First, velocity is
not proportional to the canonical momentum, see Eq. (\ref{uf4.5}). Second, in interacting theory we necessarily have
the Pirani condition $J^{\mu\nu}{\cal P}_\nu=0$ on the place of Frenkel condition $J^{\mu\nu}\dot x_\nu=0$. In the
Lagrangian formulation this leads to the equation $\left[\frac{\tilde T\dot x}{\sqrt{-\dot xg\dot x}}\right]\dot{}=f$,
which has the structure different from that of Frenkel and BMT, $\left[\frac{\dot x}{\sqrt{-\dot x\dot
x}}\right]\dot{}=f$. This results in extra contribution to the standard expression for the Lorentz force, $\ddot x\sim
F\dot x+O(J)$. So the complete theory implies an extra spin-orbit interaction as compared with the approximate Frenkel
and BMT equations. For instance, BMT electron in a constant magnetic field moves around a circle on the plane
orthogonal to the field. For our particle, the circular motion is perturbed by slow oscillations along the magnetic
field \cite{DPM3}.

Frenkel condition implies $\vec D=0$ in the rest frame, that is zero electric dipole moment. In contrast, the Pirani
condition (\ref{L.9}) predicts small non-vanishing electric dipole moment $\vec D\sim \vec S\times(\vec S\times\vec
E)$.
%
%

As it should be in a Lorentz-invariant theory, the speed of light $c$ represents the invariant scale in our model: if
one observer concludes that a particle has the speed $c$, all other inertial observers will make the same conclusion.
At the same time, when $\mu\ne 1$ our equations of motion necessarily involve the factor $\sqrt{-\dot xg\dot x}$
instead of the standard relativistic-contraction factor $\sqrt{-\dot x^2}$. Computing the acceleration implied by
(\ref{L.7})-(\ref{L.9}), we obtain ${\vec a}\sim\sqrt{-\dot xg\dot x}\vec f$ with ${\vec f}$ being non-singular
function as $\dot xg\dot x\rightarrow 0$. So the factor determines critical speed ${\vec v}_{cr}$ which the spinning
particle can not overcome during its evolution in external field. The critical speed is determined as a solution to
$\dot xg\dot x=0$. This surface is slightly different from the sphere $c^2-{\vec v}^2=0$. Indeed, we compute
\begin{eqnarray}\label{L.14.2}
-(\dot xg\dot x)=c^2-{\vec v}^2+4b^2(\mu-1)^2\left[\pi^2(\omega F\dot x)^2+\omega^2(\pi F\dot x)^2\right]\,.
\end{eqnarray}
As $\pi$ and $\omega$ are space-like vectors, the last term is non-negative, so $|{\vec v}_{cr}|\ge c$. Let us confirm
that this term not always vanishes as $|{\vec v}|=c$, that is critical velocity could be different from $c$. Assume the
contrary, that the last term in (\ref{L.14.2}) vanishes, then
\begin{eqnarray}\label{L.14.3}
\omega F\dot x=-\omega^0({\vec E}{\vec v})+(\vec\omega, c{\vec E}+{\vec v}\times{\vec B})=0\,, \cr \pi F\dot
x=-\pi^0({\vec E}{\vec v})+(\vec\pi, c{\vec E}+{\vec v}\times{\vec B})=0\,.
\end{eqnarray}
This implies (see the notation (\ref{m.10}) and (\ref{L.14.6})) $c({\vec D}{\vec E})+({\vec D}, {\vec v}\times{\vec
B})=0$. Consider the case ${\vec B}=0$, then it should be $({\vec D}{\vec E})=0$. On other hand, for the homogeneous
field the quantity $J^{\mu\nu}F_{\mu\nu}=2\left[({\vec D}{\vec E})+2({\vec S}{\vec B})\right]=2({\vec D}{\vec E})$ is a
constant of motion \cite{DPM3}. Let us take the initial conditions for spin such, that $({\vec D}{\vec E})\ne 0$. Then
critical speed of our particle in this field will be different from the speed of light. Similar conclusion has been
made by Hanson and Regge with respect to their relativistic spherical top \cite{hanson1974relativistic}.

Detailed discussion of the ultra-relativistic motion is presented in \cite{AAD-WGR:2014}

\section{Acknowledgments}
This work has been supported by the Brazilian foundation CNPq (Conselho Nacional de Desenvolvimento Científico  e
Tecnológico - Brasil).

\end{document}